\documentclass{ws-procs9x6}

\begin{document}

\title{High energy scattering in QCD: dipole approach with Pomeron loops}

\author{Eugene Levin \footnote{\uppercase{W}ork partially
supported by  the  \uppercase{I}srael \uppercase{S}cience \uppercase{F}oundation, 
founded by the \uppercase{I}sraeli \uppercase{A}cademy of \uppercase{S}cience and 
\uppercase{H}umanities. }}

\address{Department of Particle Physics, School of Physics and Astronomy, \\ 
Raymond and Beverly Sackler Faculty of Exact Science\\
Tel Aviv University, Tel Aviv, 69978,Israel\\ 
E-mail: leving@post.tau.ac.il, levin@mail.desy.de, elevin@quark.phy.bnl.gov\\
~\\
Talk given at ``Gribov-75 Memorial Workshop on Quarks, Hadrons, and Strong Interactions", 
  May 22-24, 2005, Budapest, Hungary}

\maketitle

\abstracts{
In this talk we discuss the BFKL Pomeron Calculus and its interrelation with the 
colour dipole approach. The two key problems we consider are the  
probabilistic interpretation of the BFKL Pomeron Calculus and the possible 
scenario for the asymptotic behaviour of the scattering amplitude at high energy 
in QCD.}

\section{Introduction}
\label{sec:int}
The simplest approach that we can propose for  high energy interaction is 
based \cite{GLR,MUQI} on the BFKL Pomeron \cite{BFKL} and reggeon-like diagram technique
for the BFKL Pomeron interactions \cite{BART,BRN,NP,BLV}. This technique which is a 
generalization of Gribov Reggeon Calculus \cite{GRC},  can be written 
in the elegant form of the functional integral (see Ref. \cite{BRN} and the next 
section); and it is a challenge to solve this theory in QCD  finding the high energy 
asymptotic behaviour.  However, even this simple approach has not been solved during 
three decades of attempts by the high energy community. This failure stimulates a search 
for deeper understanding of physics which is behind the BFKL Pomeron Calculus. In 
particular, at the end of the Reggeon era it was understood \cite{GRPO,LEPO,BOPO}  that 
the Reggeon Calculus 
can be reduced to a Markov process \cite{GARD} for the probability of finding a given 
number of Pomerons at fixed rapidity $Y$.  Such an interpretation, if it would be 
reasonable in QCD,  can be useful,  
 since it allows us to use powerful methods of statistical 
physics in our search for the solution.

The goal of this talk is to consider two main problems: the probabilistic 
interpretation of the  BFKL Pomeron Calculus based on the idea that colour dipoles 
are the correct degrees of freedom at high energy QCD \cite{MUCD} and the possible 
solution for the scattering amplitude at high energy.
Colourless dipoles play two different roles in our approach. First, they are 
partons (`wee' partons) for the BFKL Pomeron. This role is not related to the large 
$N_c$ approximation and, in principle, we can always speak about probability to 
find a definite number of dipoles instead of defining the probability to find a 
number of the BFKL Pomerons. The second role of the colour dipoles is that at high 
energies we can  interpret the vertices of Pomeron merging and splitting in terms 
of the probability for two dipoles to annihilate in one dipole and of the probability for the
decay of one dipole into two. It was shown in Ref. \cite{MUCD} that $P 
\rightarrow 2P$ splitting can be described as the process of the dipole decay into 
two dipoles. However, the relation between the Pomeron merging ($2 P \rightarrow 
P$) and the process of annihilation of two dipoles into one dipole is not so 
obvious 
and it will be discussed here.

 Our presentation is based on Refs.\cite{L1,L2,L3,L4,L5}
and we would like to thank Michael Lublinsky and Alex Prygarin for their 
contributions and for a great pleasure to work with them.

The outline of the talk looks as follows. In the next section we will discuss the 
BFKL Pomeron Calculus in the elegant form of the functional integral,  suggested by 
M. Braun about five years ago \cite{BRN}. We  will show that the  intensive 
recent work on this subject \cite{IT,MSW,L3} confirms the BFKL Pomeron Calculus in 
spite of the fact that these attempts were based on slightly different but not more 
general  assumptions. 

In the third section the general approach based on the generating functional will 
be discussed. The set of  equations for the amplitude of $n$-dipole interaction 
with the target will be obtained and the interrelations between vertices of the 
Pomeron interactions and the microscopic dipole processes will be considered.

The fourth section is devoted to the toy model which simplifies the QCD interaction 
and allows us to see the main properties of high energy amplitude. In particular, 
we are going to discuss the solution of the equations for the asymptotic behaviour 
of the scattering amplitude at high energies.

In the fifth section we will consider a general case of  QCD and will discuss the 
probabilistic interpretation as well as the scenario for the solution using the 
experience with the toy model.

In conclusion we are going to compare our approach with other approaches on the 
market. 

\section{The BFKL Pomeron Calculus}
The main ingredient of this calculus is the Green function  of the BFKL 
Pomeron describing the propagation of a pair of gluons from rapidity $y'$ and 
points 
$x'_1$ and $x'_2$ to rapidity $y$ and points $x_1$ and $x_2$ \footnote{Coordinates 
$x_i$ here are two dimensional vectors and, strictly speaking, should be denoted by 
$\vec{x}_i$ or $\bf{x}_i$. However, we will use the notation $x_i$ hoping that it will 
not cause difficulties in understanding.}. Since the Pomeron does not carry colour 
in the $t$-channel, we can treat initial and final coordinates as coordinates of quark 
and antiquark in a colourless dipole. This Green function is well known\cite{LI} 
and has a form:
\begin{equation} \label{BFKLGF}
G(x_1,x_2;y | x'_1,x'_2;y')\,\,=\,\,\Theta(y - y')\times
\end{equation}
$$
\times\,\sum^{\infty}_{n=-\infty} \,\int\frac{d \nu}{2 \pi 
i}\,d^2 x_0\,e^{\omega(n,\nu) (y - y')}\,\lambda(n,\nu)\, 
E(x_1,x_2;x_0|\nu)\,E^*(x'_1,x'_2;x_0|\nu) 
$$
where vertices $E$ are given by

\begin{equation} \label{BFKLE}
E(x_1,x_2;x_0|\nu)\,=\,
\left(\frac{x_{12}}{x_{10}\,x_{20}} 
\right)^h\,\left(\frac{x^*_{12}}{x^*_{10}\,x^*_{20}} \right)^{\tilde{h}} ;
\end{equation}
$x_{ik} = x_i - x_k $, $x_i = x_{i,x} + i x_{i,y}$ \footnote{$x_{i,x}$ and  
$x_{i,y}$  are components of the two dimensional vector $x_i$ on $x$-axis and $y$- 
axis} ,$ x^*_i = x_{i,x} + i x_{i,y} $ ; $h = (1 - n)/2 + i\nu$ and $\tilde{h} = 1 
- h^*$. The energy levels $\omega(n,\nu)$  are the BFKL eigenvalues
\begin{equation} \label{BFKLOM}
\omega(n,\nu)\,=\,\bar{\alpha}_S \left( \psi(1) - Re \psi(\frac{|n| + 1}{2} + i 
\nu) 
\right)
\end{equation}
where $\Psi(z) = d \ln \Gamma(z)/d z$ and $\Gamma(z)$ is the Euler gamma function. 
 $\bar{\alpha}_S = \alpha_S N_c/\pi$ and finally
\begin{equation} \label{BFKLLA}
\lambda(n,\nu)\,=\frac{1}{[ ( n + 1)^2 + 4 \nu^2] [(n - 1)^2 + 4 \nu^2]}.
\end{equation}

The interaction between Pomerons is described by the triple Pomeron vertex which 
can be written in the coordinate representation \cite{BRN} for the 
following process: two gluons with coordinates $x_1$ and $x_3$ at rapidity $y_1$  
decay into two gluons pairs with coordinates $x"_1$ and $x"_2$ at rapidity $y_2$  
and  $x"_3$ and 
$x"_4$ at rapidity $y_3$ due to the Pomeron splitting at rapidity $y$. It looks 
like
\begin{equation} \label{BFKL3P}
\frac{\pi\,\bar{\alpha}^2_S}{N_c}\,\int\,\frac{d^2 x'_1\,d^2\,x_2 \,d^2 
x_3}{x^2_{12}\,x^2_{23}\,x^2_{13}}\,\left(G(x_1,x_3;y_1|x'_1,x'_3; y) 
\,\,L^{\!\!\!\!\!\!\!\leftarrow}_{1',3'} \right) \cdot
\end{equation}
$$ 
G(x'_1,x'_2;y|x"_1,x"_2;y_2)\,G(x'_2,x'_3;y|x"_3,x"_4;y_2)
$$
where 
\begin{equation} \label{BFKLL}
L^{\!\!\!\!\!\!\!\leftarrow}_{1',3'}\,\,=\,\,r'^4_{13}\,p^2_{1'}\,p^2_{3'}\,\,
\,\mbox{with}\,\,p^2\,=\,-\,\nabla^2
\end{equation}
and the arrow shows the direction of action of the operator $L$.
For the inverse process of merging of two Pomerons into one we have 
$$
\frac{\pi\,\bar{\alpha}^2_S}{N_c}\,\int\,\frac{d^2 x'_1\,d^2\,x_2 \,d^2
x_3}{x^2_{12}\,x^2_{23}\,x^2_{13}}\,
G(x"_1,x"_2;y_2|x'_1,x'_2;y)\,G(x"_3,x"_4;y_2|x'_2,x'_3;y)\,\cdot
$$
\begin{equation} \label{BFKLP3}
\left( L^{\!\!\!\!\!\rightarrow}_{1',3'}\,
G(x'_1,x'_3; y|x_1,x_3;y_1) \right) 
\end{equation}

The theory with the interaction given by Eq.~(\ref{BFKL3P}) and  Eq.~(\ref{BFKLP3})
can be written through the functional integral \cite{BRN}
\begin{equation} \label{BFKLFI}
Z[\Phi, \Phi^+]\,\,=\,\,\int \,\,D \Phi\,D\Phi^+\,e^S \,\,\,\mbox{with}\,S \,=\,S_0 
\,+\,S_I\,+\,S_E
\end{equation}
where $S_0$ describes free Pomerons, $S_I$ corresponds to their mutual interaction 
while $S_E$ relates to he interaction with the external sources (target and 
projectile). From Eq.~(\ref{BFKL3P}) and  Eq.~(\ref{BFKLP3}) it is clear that
\begin{equation} \label{S0}
S_0\,=\,\int\,d y \,d y'\,d^2 x_1\, d^2 x_2\,d^2 x'_1\, d^2 
x'_2\,
\end{equation}
$$
\Phi^+(x_1,x_2;y)\,
G^{-1}(x_1,x_2;y|x'_1,x'_2;y')\,\Phi(x'_1,x'_2;y')
$$
\begin{equation} \label{SI}
S_I\,=\,\frac{\pi \tilde{\alpha}^2_S}{N_c}\,\int \,d y\,\int \,\frac{d^2 x_1\,d^2 
x_2\,d^2 x_3}{x^2_{12}\,x^2_{23}\,x^2_{13}}\,\cdot
\end{equation}
$$
\cdot
\{ 
\left(L^{\!\!\!\!\!\rightarrow}_{1,3}\,\Phi(x_1,x_3;y) \right)\,\cdot\, 
\Phi^+(x_1,x_2;y)\,\Phi^+(x_2,x_3)\,\,+\,\,h.c. \}
$$
For $S_E$ we have local interaction both in rapidity and in coordinates, namely,
 \begin{equation} \label{SE}
S_E\,=\,-\,\int \,dy\,d^2 x_1\,d^2 x_2\,
\end{equation}
$$
\{ 
\Phi(x_1,x_2;y)\,\tau_{pr}(x_1,x_2;y)\,\,+\,\,\Phi^+(x_1,x_2;y)\,\tau_{tar}(x_1,x_2;y)
$$
where $\tau_{pr}$ ($\tau_{tar}$ stands for the projectile and target, respectively. 
The form of function $\tau$  depends on the non-perturbative input in our problem 
and for the case of nucleus target they are written in Ref. \cite{BRN}.

For the case of the projectile being a dipole that scatters off a  nucleus
the scattering amplitude has the form
\begin{equation} \label{T}
T(Y,x_1,x_2)\,\,=\,\,\frac{4\,\pi^2\,\bar{\alpha}_S}{N_c}\, \frac{\int\,\,D \Phi\, 
D\Phi^+\,\Phi(x_1,x_2;Y)\,e^S}{ \int\,\,D 
\Phi\, D \Phi^+\,\,e^S|_{S_E =0}}
\end{equation}
where the extra $\alpha_S$ comes from our normalization.

For further presentation we need some properties of the BFKL Green function 
\cite{LI}:

1.  Generally,

\begin{equation}
G^{-1}(x_1,x_2;y| x'_1,x'_2;y')\,\, \,=\,p^2_1\,p^2_2\,\left( 
\frac{\partial}{\partial y} + H \right) \,\,=\,\,\left(
\frac{\partial}{\partial y} + H^+ \right)\,p^2_1\,p^2_2; \label{G1}
\end{equation}

\begin{equation}
H f(x_1,x_2;y) \,\,=
\end{equation}
$$
\,\,\frac{\bar{\alpha}_S}{2 \pi}\,\int\,\frac{d^2 
x_3\,x^2_{12}}{x^2_{23}\,x^2_{13}}\,\left( 
f(x_1,x_2;y)\,-\,f(x_1,x_3;y)\,-\,f(x_3,x_2;y) \right) ; \label{G2}
$$

2. The initial Green function ($G_0$) is
 equal to 
\begin{equation} \label{G0}
G_0(x_1,x_2;y| x'_1,x'_2;y')\,\,=\,\, \frac{\pi^2}{2} \ln 
\frac{x^2_{1,1'}\,x^2_{2,2'}}{x^2_{1,2'}\,x^2_{1',2}}  \,\ln 
\frac{x^2_{1,1'}\,x^2_{2,2'}}{x^2_{1,2}\,x^2_{1',2'}}
\end{equation}
 
3. It should be stressed that
\begin{equation} \label{G01}
\nabla^2_1 \,\nabla^2_2 \,G_0(x_1,x_2;y| x'_1,x'_2;y')\,\,=
\end{equation}
$$
\,\,\delta^{(2)}( x_1 - 
x'_1)\,\delta^{(2)}( x_2 -
x'_2)\,+\,\delta^{(2)}( x_1 -
x'_2)\,\delta^{(2)}( x_2 -
x'_1)\,
$$

4. In the sum of Eq.~(\ref{BFKLGF}) only the term with $n=0$ is essential for the high 
energy asymptotic behaviour since all $\omega(n,\nu) $ with $n   \geq 1$  are 
negative and, therefore, 
lead to contributions that decrease with energy. Taking into account only the 
first 
term, one can see that $G$ is the eigenfunction of operator $L_{13}$, namely
\begin{eqnarray}
L_{13}\,G_0(x_1,x_2;y| x'_1,x'_2;y')\,\,&=&\,\,\frac{1}{\lambda(0,\nu)} 
\,G_0(x_1,x_2;y| x'_1,x'_2;y')\,\nonumber \\
 & & \,\approx\,\,G_0(x_1,x_2;y| x'_1,x'_2;y') ; \label{L13}
\end{eqnarray}
The last equation holds only approximately in the region where $\nu \,\ll\,1$, but 
this is the most interesting region which is responsible for the high energy 
asymptotic behaviour of the scattering amplitude.

\begin{figure}[ht]
\centerline{\epsfxsize=3.2in\epsfbox{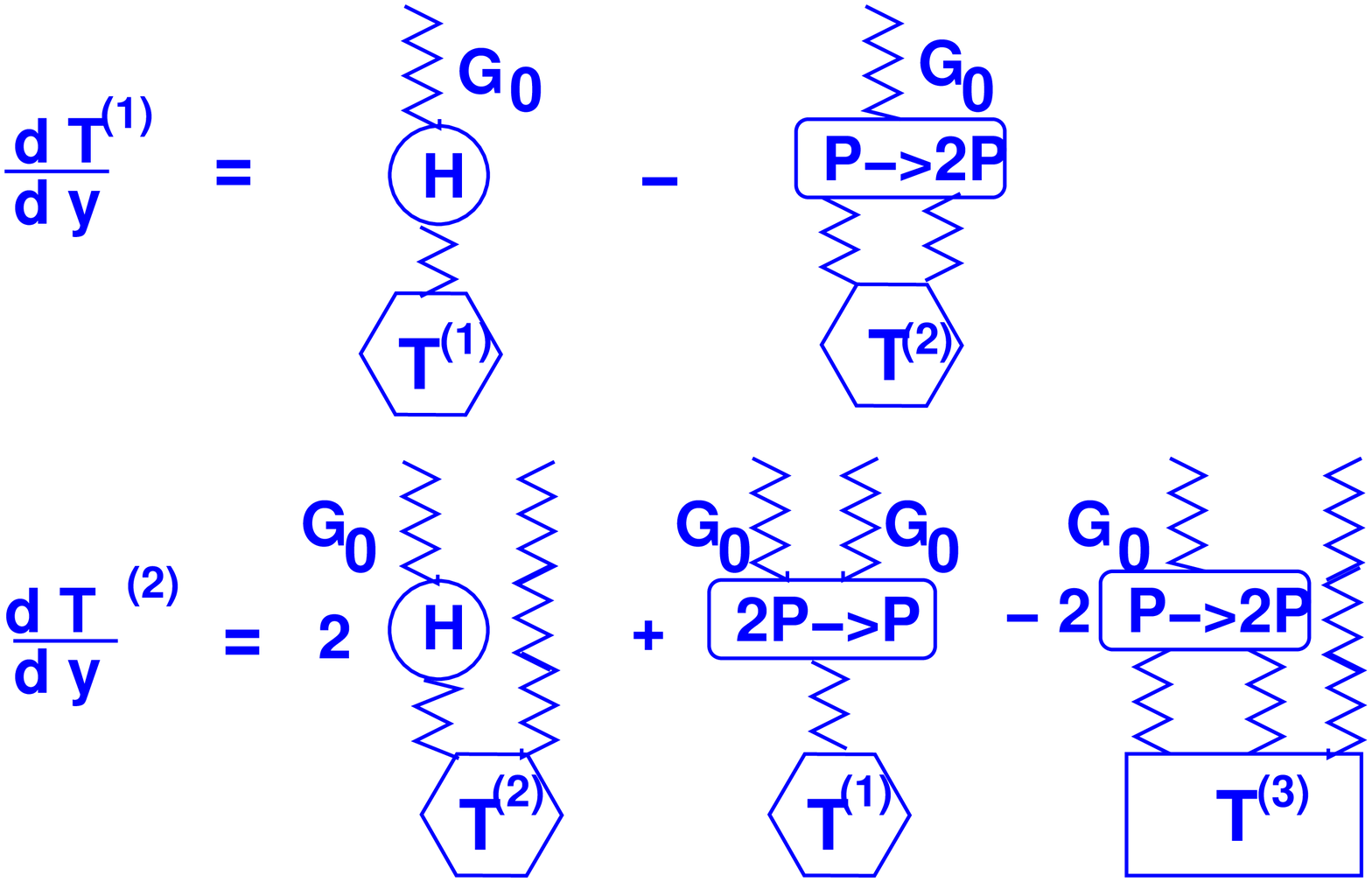}}   
\caption{The graphic form of equations for the multi dipole amplitude. 
\label{pomeq}}
\end{figure}

Using  Eq.~(\ref{BFKL3P}),  Eq.~(\ref{BFKLP3}) and Eq.~(\ref{G2})
 we can easily obtain the chain equation for multi-dipole amplitude $T^{(n)}$  
noticing that every dipole interacts only with one Pomeron (see Eq.~(\ref{T})). 
The 
equation is shown in Fig. ~(\ref{pomeq}) and it has the following form
\begin{equation} \label{T1}
\frac{\partial T^{(1)}(y;x_1,x_2)}{\partial y}\,\,=\,\, \frac{\bar{\alpha}_S}{2 
\,\pi}\,\int\,d^z\,K(x_1,x_2;z)\,
\end{equation}
$$
\left( 
T^{(1)}(y;x_1,z)\,+\,T^{(1)}(y;z,x_2)\, -\, 
T^{(1)}(y;x_1,x_2\,-\,T^{(2)}(y;x_1,z;z,x_2)\right) 
$$
\begin{equation}  \label{T2}
\frac{\partial T^{(2)}(y;x_1,x_2;x_3,x_4)}{\partial y} \,\,=\,\,  
\frac{\bar{\alpha}_S}{2
\,\pi}\int\,d^z\,K(x_1,x_2;z)\,
\end{equation}
$$
\left(  T^{(2)}(y;x_1,z;x_3,x_4) 
\,
+\,T^{(2)}(y;z,x_2;x_3,x_4) \,-\,
\,T^{(2)}(y;x_1,x_2;x_3,x_4)\, - \right.
$$
$$
\left.  
-\,T^{(3)}(y;x_1,z;z,x_2;x_3,x_4)\,+\,\{x_1 \to x_3, x_2 \to x_4\} \right)\,+
$$
$$
+\,
(\frac{4 \pi^2\bar{\alpha}_S}{N_c})^2  \frac{\bar{\alpha}_S}{2
\,\pi} \int \Gamma_{2 \to 1}(x_1,x_2;x_3,x_4| x',x")\,\nabla^2_{x'}\nabla^2_{x"} 
T^{(1)}(y;x',x"); 
$$
Deriving Eq.~(\ref{T1}) and Eq.~(\ref{T2}) we use Eq.~(\ref{G0}) and 
Eq.~(\ref{G01}) as well as the normalization condition (see Eq.~(\ref{T})) for 
the 
scattering amplitude. These two equations are the same as in Ref. \cite{IST}. 
This 
 shows that in the  papers \cite{IT,IST}, actually,  the same approach is 
developed as in the BFKL 
Pomeron Calculus (much later, of course), in spite of the fact 
that the authors think that they are doing something more general.   

Assuming that $T^{(2)}= T^{(1)}\,T^{(1)}$, we obtain the Balitsky-Kovchegov 
equation 
\cite{B,K}. We can do this only if we can argue why the Pomeron splitting is more 
important than the Pomeron merging. For example this assumption is reasonable 
for the 
scattering of the dipole with the nucleus target.  Generally speaking, the 
splitting 
and merging have the same order in $\alpha_S$ ( see Eq.~(\ref{BFKL3P}) and  
 Eq.~(\ref{BFKLP3}). In  Eq.~(\ref{T1}) and  Eq.~(\ref{T2}) these two processes 
look 
like having a different order of magnitude in $\alpha_S$ but this fact does not 
interrelate with any physics and reflect only our normalization. However, we will 
see that for a  probabilistic interpretation the correct normalization is very 
important. 

Kernel $K(x,y|z)$ is defined as
\begin{equation} \label{K}
K(x,y|z)\,\,=\,\,\frac{(x - y)^2}{(x - z)^2\,(z - y)^2} .
\end{equation}

Function $\Gamma_{2\to 1}$ is equal to
\begin{equation} \label{V21}
\Gamma_{2 \to 1}(x_1,y_1;x_2,y_2|x,y)\,\,=
\end{equation}
$$
=\,\,
\int\,d^2 z K(x,y;z)\,\,G_0(x_1,y_1|x,z)\,G_0(x_1,y_1|z,y)
$$
where $G_0$ is given by Eq.~(\ref{G0}) and we use the notation $(x,y)$ for a 
colourless pair of gluon (colour dipole).

\section{Generating functional and probabilistic interpretation}

In this section we discuss the main equations of the BFKL Pomeron Calculus in the 
formalism of the generating functional,  which we consider as the most 
appropriate technique for the probabilistic interpretation of this approach top 
high energy scattering in QCD.

To begin with, let us write down the definition of the generating functional 
\cite{MUCD}
\begin{equation} \label{Z}
Z\left(Y\,-\,Y_0;\,[u] \right)\,\,\equiv\,\,
\end{equation}
$$
\equiv\,\,\sum_{n=1}\,\int\,\,
P_n\left(Y\,-\,Y_0;\,x_1, y_1; \dots ; x_i, y_i; \dots ;x_n, y_n
 \right) \,\,
\prod^{n}_{i=1}\,u(x_i, y_i) \,d^2\,x_i\,d^2\,y_i  \, ,
$$
where $u(x_i, y_i) \equiv u_i $ is an arbitrary function of $x_i$ and $y_i$.
The coordinates $(x_i,y_i)$ describe the colorless pair of gluons or a dipole. 
$P_n$ is a probability density to find $n$ dipoles with the size $x_i - y_i$, 
and with the impact
parameter $(x_i + y_i)/2$. It follows from the physical meaning of $P_n$ and 
the definition in Eq.~(\ref{Z}) directly
that the functional (Eq.~(\ref{Z})) obeys the condition
\begin{equation} \label{ZIN1}
Z\left(Y\,-\,Y_0;\,[u=1]\right)\,\,=\,\,1\,.
\end{equation}
The physical meaning of this equation  is that the sum over
all probabilities is equal to unity.

Introducing vertices for the dipole process: $1 \to 2$ ($V_{1 \to 2}( x,y \to 
x_1,y_1 + x_2,y_2)$),  $2 \to 1$ ($V_{2 \to 1}( x_1,y_1 + x_2,y_2 \to x,y)$)
and $2 \to 3$ ($V_{2 \to 3}(x_1,y_1 + x_2, y_2 \to x'_1,y'_1 + x'_2, y'_2  + 
x'_3,y'_3)$ we can write a typical birth-death equation in the form
\begin{equation} \label{P}
\frac{\partial \,P_n(\dots,x_i,y_i; \dots)}{\partial Y}\,\,\,=
\end{equation}
\begin{eqnarray}
 &=&\,\, \sum_{i} \  V_{1 \to
2}
\bigotimes \left(P_{n-1}(\dots,x_i,y_i; \dots)\, -\,P_n (\dots,x_i,y_i; 
\dots)\right) \label{P1} \\
 & & \,\, \sum_{i > k} V_{2 \to 1}\,\left( P_{n+1}(; x_i,y_i; \dots ; x_k,y_k; ) 
\,- \, P_n(;x_i,y_i;) \right)  \label{P2} \\
 & & \,\, \sum_{i > k}\, V_{2 \to 3}\bigotimes  \left( P_{n 
-1}(;x_i,y_i; ; x_k,y_k;)\,-\,P_n(;x_i,y_i; ; x_k,y_k;)\right) \label{P3}
\end{eqnarray}
 Multiplying this equation  by the product $\prod^n_{i=1}\,u_i$
and integrating over all $x_i$ and $y_i$,  we obtain the
following linear equation for the generating functional:
\begin{equation}\label{ZEQ}
\frac{\partial \,Z\,[\,u\,]}{
\partial \,Y}\,\,= \,\,\chi\,[\,u\,]\,\,Z\,[\,u\,]
\end{equation}
with
\begin{equation} \label{chi}
\chi[u]\,\,=\,\, -\,\int\,d^4\,q d^4 q_1\,d^4 q_2 \,\,
\end{equation}
\begin{eqnarray}
& & \left(  V_{1\,\rightarrow \,2}\left( q \to 
q_1 + q_2 \right)\, \left( - u(q) \,+\,u(q_1) \,u(q_2)\,\right) 
\,\frac{\delta}{\delta u(q)}\,- \right. 
\label{VE12} \\
 & &\left. - V_{2\,\rightarrow \,1}\left(  q_1 + q_2 \to q  \right)\,
\left( u(q_1)  \,u(q_2) \,-\, u(q) \right) \,\,\frac{1}{2} \,\frac{\delta^2}{\delta 
u(q_1)\,\delta 
u(q_2)} \right)\,.
\label{VE21} 
\end{eqnarray}

 Trying to make our presentation more transparent,   we omitted  in  
Eq.~(\ref{chi})    
the
term that corresponding to   the $ 2 \to 3$ transition (see Ref. \cite{L3} for 
full 
presentation).

Eq.~(\ref{ZEQ})  is a typical diffusion equation or Fokker-Planck equation 
\cite{GARD},   with a
diffusion
coefficient depending on $u$. This is the master equation of our approach, and 
the goal of this
talk is to find the correspondence of this equation with the BFKL Pomeron 
Calculus 
and the asymptotic
solution to this equation. In spite of the fact that this is a
functional equation we intuitively feel that this equation could be useful since 
we can develop a direct method for its solution and, on the other hand, there 
exist many studies of such
an equation
in the literature ( see for example Ref. \cite{GARD}) as well as some physical 
realizations in
statistical physics. The intimate relation between the Fokker-Planck equation
 and 
the high energy
asymptotic was first pointed out  by Weigert \cite{WE} in JIMWLK approach 
\cite{JIMWLK},  and has 
been discussed in
Refs.
\cite{BIW,IT,MSW}.

The scattering amplitude can be defined as a functional \cite{K,L2}
\begin{eqnarray} 
& N\left(Y;[\gamma_i] \right)\,= & \label{N}\\
 &=\,- \sum^{\infty}_{n =1} \int 
\gamma_n(x_1,y_1;\dots;x_n,y_n;Y_0)\prod^n_{i=1}\frac{\delta}{\delta 
u_i}Z\left(Y,[u_i]\right)|_{u_i=1}\,d^2 x_i\,d^2 y_i& \nonumber\\
 &= \,-\,\sum^{\infty}_{n =1} (-1)^n \int 
\gamma_n(x_1,y_1;\dots;x_n,y_n;Y_0)\rho(x_1,y_1;;x_n,y_n;Y -Y_0)& \nonumber \, .
\end{eqnarray}

The physical meaning of functions $\gamma_n$ is the imaginary part of the amplitude 
of the interaction of $n$ dipoles with the target at low energies. All these functions 
should be taken from the non-pertubative QCD input. However, in Refs. 
\cite{L1,L2,L3} 
it was shown that we can introduce the amplitude of interaction of $n$ dipoles
($\gamma_n(x_1,y_1;\dots;x_n,y_n;Y)$ at 
high energies (large values of rapidity $Y$) and Eq. ~(\ref{Z}), Eq. ~(\ref{ZEQ})
and Eq. ~(\ref{N}) can be rewritten as a chain set of equation for 
$\gamma_n(x_1,y_1;\dots;x_n,y_n;Y)$.  The equation has the form\footnote{This 
equation is Eq.~(2.19) of  Ref. \cite{L3} but, hopefully, without misprints , 
part of which has been noticed in Ref. \cite{IST}.}
\begin{eqnarray} \label{N5}
{}&{} & \frac{\partial\, \gamma_n\left( r_1,b_1,\dots, r_n,b_n 
\right)}{\partial\,Y}\,\,\,\,=  2\,\sum_{i=1}^n\,\int\,d^4q'\,d^4q \,
V_{1\,\rightarrow \,2} (q_i;\,q\,q')\,
\gamma_n\left(  \dots q'\dots \right) \,\nonumber
\end{eqnarray}
\begin{eqnarray}
&-&\,   \sum_{i=1}^n \,\int d^4q_1'\,d^4q_2'\,
V_{1 \,\rightarrow\,2} (q_i;\,q_1'\,q_2')\,\gamma_n\left(  \dots, q_i\dots 
\right)\,-\sum^{n - 1}_{i =1} \int d^4q\,d^4q' \\
& & V_{1\,\rightarrow \,2} (q_i;\,q\,q')\,\gamma_{n+1}\left(  \dots q \dots 
q'\right)
-\,\sum_{i > j}^n\,\int  d^4\,q\,
V_{2\,\rightarrow\,1}\left( q_i\,q_j ;\,q \right) \nonumber \\
&\cdot&\gamma_{n-1}\left(
 q_i \dots  q_j \dots q\right)
\,+\, 2\,\sum_{i=1}^n \int d^4 q\, d^4 q' V_{2
\rightarrow 1} \left( q q_i;q'\right) \, \gamma_{n-1}\left( \dots q_i \dots
q\right)
\,+ \nonumber \\
&+&
\,\sum^n_{i,k,i > k}\,   \int d^4q\,
V_{2\rightarrow 1} \left( q_i q_k; q \right)\,\gamma_n\left(  \dots
q_i\dots q_k \dots \right)
\nonumber
\end{eqnarray}

Comparing this equation for $\gamma_1 \equiv T^{(1)}$ and $\gamma_2 \equiv T^{(2)}$ 
one can see that 
\begin{eqnarray}
V_{1 \to 2} &=&\frac{\bar{\alpha}_S}{2\,\pi} \Gamma_{1 \to 
2}\,\,=\,\,\frac{\bar{\alpha}_S}{2\,\pi}\,K\left(x,y;z\right)\,; \label{GA1}\\
V_{2 \to 1}\,&=&\, (\frac{4 \pi^2\bar{\alpha}_S}{N_c})^2  \frac{\bar{\alpha}_S}{2
\,\pi }\, \left(-\,\Gamma_{2 \to 1}(x_1,y_1 + x_2,y_2 \to x,y) \,+ \right. 
\nonumber \\
&+& \left.
 \,
\,\int 
d^2x\,d^2y\,\Gamma_{2 \to 1}(x_1,y_1 + x_2,y_2 \to x,y) \,\cdot \right. \nonumber 
\\
 & \cdot& \left. \left(\delta^{(2)}(x_1 - x)\delta^{(2)}(y_1 - y)\,+\,\delta^{(2)}(x_2 - 
x)\delta^{(2)}(y_2
- y)\right) \right)\label{GA2}
\end{eqnarray}
with $\Gamma_{2 \to 1}$ is given by Eq.~(\ref{V21}).
\section{A toy model: Pomeron interaction and probabilistic interpretation}
In this section we consider the simple toy model in which the probabilities to find 
$n$-dipoles do not depend on the size of dipoles \cite{MUCD,L1,L3,L4}. In this 
model the master equation (\ref{ZEQ}) has a simple form
\begin{equation} \label{TMZ}
\frac{\partial Z}{\partial Y}\,=\,- \Gamma(1 \to 2)\,u (1 - u) 
\,\frac{\partial Z}{\partial u}+\,\Gamma(2 \to 
2)\,u (1 - u)\,\frac{\partial^2 Z}{(\partial u)^2}
\end{equation} 
and this equation generates the Pomeron splitting $G_{P \to 2P}= - \Gamma(1 \to 
2)$, Pomerons merging $G_{2P \to P}=  \Gamma(2 \to 1)$  and also the two Pomeron 
scattering $G_{2P \to 2P}\,=\,- \Gamma(2 \to 1)$.  It is easy to see that 
neglecting the $u^2 \partial^2 Z/(\partial u)^2$ term in Eq.~(\ref{TMZ}), 
we cannot 
provide a correct sign for Pomerons merging $G_{2P \to P}$.

Eq.~(\ref{TMZ}) is the diffusion with the $u$ dependence in the diffusion 
coefficient.  
 For $u<1$ the diffusion coefficient is positive and the equation has a reasonable 
solution. If $u >1$, the sign of this coefficient changes and the equation gives a 
solution which increases with $Y$ and $Z(Y)$  cannot be treated as the
generating function for the probabilities to find $n$ dipoles (Pomerons) (see Refs. 
\cite{GRPO,BOPO,L4} for details). We can see the same features in the asymptotic 
solution that is the solution to Eq.~(\ref{TMZ}) with the l.h.s. equal to zero.
It is easy to see that this solution has the form
\begin{equation}
Z(u;Y \to \infty) \,\,=\,\,e^{-\,\kappa\,(1 - u)}\,\,\,\,\mbox{with}\,\,\, 
\kappa\,=\,
\frac{\Gamma(1 \to 2)}{\Gamma(2 \to 1)}\,=\,\frac{2 
\,N^2_c}{\bar{\alpha}^2_S}\,\gg\,1
\end{equation}
One can see that for negative $\kappa$ this solution leads to $Z>1$ for $u <1$. 
This shows that we cannot give a probabilistic interpretation for such a solution. 

Using the asymptotic solution and finding the typical $u$ at high energies from 
the unitarity constraints~\cite{L4,IM,KOLE} we can   
determine the asymptotic limit for
the high energy amplitude. It turns out that
\begin{equation} \label{TMNAS}
N(Y \to \infty)\,\,=\,\, 1 \,\,-\,\,e^{- \kappa}
\end{equation}

It was shown in Ref. \cite{L4} that we can search for the correction to the 
solution of Eq.~(\ref{TMNAS}) in the form 
\begin{equation} \label{TMNAS1} 
Z(u; Y)\,\,=\,\,Z(u;Y \to \infty)\,\,+\,\,\Delta Z(u; Y) 
\end{equation}
Asssuming\,\,$\Delta Z(u; Y)(u;Y \to \infty)\,\,\gg\,\,Z(u;Y \to \infty)$
we can  write
for $\Delta Z(u; Y)$ the linear equation\cite{L4} and the solution of 
this equation 
 decreases with energy.

It is interesting to notice that this asymptotic amplitude does not show a black 
disc behaviour at high energies. The behaviour of the scattering amplitude, given 
by 
Eq.~(\ref{TMNAS}), corresponds to the so called gray disc behaviour.  

\section{Probabilistic interpretation in QCD}
Eq. (\ref{GA1}) has a very simple physical meaning describing the Pomeron splitting 
as the decay process of one dipole into two dipoles. It turns out that the vertices 
for $2P \to 3P$ and $2P \to 2P$ can be easily understood as a dipole `swing'. What 
we mean is that   with some probability two quarks of
a pair  of dipoles can exchange their antiquarks to form another pair  of dipoles 
\cite{L3}.
Naturally, this process has an $1/N_c^2$ suppression and it correctly reproduces the
splitting and rescattering of two Pomerons that has been explicitly calculated from 
the diagrams \cite{BLV}.

Eq.~(\ref{GA2})\footnote{This equation is quite different from the equation which 
is obtained in Ref. \cite{L3} (see also Ref. \cite{IST}). The main difference stems 
from the  correct use of the BFKL Pomeron Calculus for determining this vertex 
while in Refs. \cite{L3,IST} the Born diagram was used which does not and cannot 
give a correct expression.}
 is more difficult to view as the vertex for the 
transition of two  dipoles into one. Indeed, the integral over  coordinates of the 
produced dipole  is positive, namely $\int V(2 \to 1) d^2 x d^2 y =  \int \Gamma(2 
\to 1) d^2 x d^2 y$. However, Eq.~(\ref{GA2}) leads, generally speaking, to a 
negative vertex in some regions of the phase space. The detailed study of this 
vertex will be published soon \cite{L5}. Here we want to point out that the key 
problem is not in the probabilistic interpretation of the microscopic process of 
two dipoles to one dipole transition but the fact that a negative vertex $V(2 \to 
1)$ means that in some kinematic region we have a negative diffusion coefficient 
which results in a solution that increases at large values of rapidity $Y$.
To save such a theory, we have to introduce other Pomeron interactions like $2P \to 
3P$ and/or $2P \to 2P$ transitions.

In Ref. \cite{L3,L4} attempts were made to deal with such Pomeron interactions. 
It turns out that at large values of $Y$ the $2 \to 1$ process contribute in a very 
limited part of the kinematic region with a very specific function $u(x,y)$ . In this 
particular region the vertex $V(2 \to 1)$ is positive \cite{L5}. Therefore, we 
could use the probabilistic interpretation but we need to study this process better 
and deeper to obtain the final result.

\begin{figure}[ht]
\centerline{\epsfxsize=3.0in\epsfbox{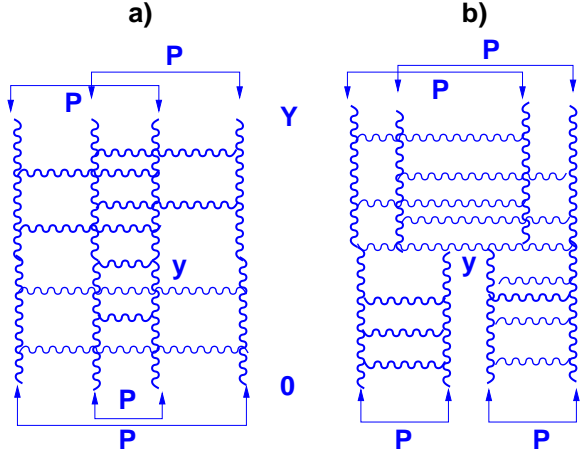}}
\caption{The diagrams for $2P \to 2P$ process. Fig.~(\ref{feymdi})-a shows the 
diagram which is of the order $\bar{\alpha}_S/N^2_c$ while  Fig.~(\ref{feymdi})-a 
leads to a contribution $\propto \bar{\alpha}^3_S$. \label{feymdi}}
\end{figure}

In the toy model it has been shown \cite{GRPO,BOPO} that we can generate the $2P 
\to 3P$ vertex without the process of two dipoles to one dipole transition. The 
transition of two dipoles to two or more dipoles also leads to this vertex.  
Similar ideas are developed in Ref. \cite{KOLU} in QCD. However, we need to pay a 
price: the contribution to Eq.~(\ref{ZEQ}) will be negative to give the correct sign 
for $2P \to P$ interaction. In other words, we can add to Eq.~(\ref{ZEQ}) the 
contribution
$$
\int\,\prod^2_{i=1}d^2 x_i\,d^2 y_i\,\prod^2_{i=1}d^2 x'_i\,d^2 y'_i\,
V_{2 \to 2}(x_1,y_1 + x_2,y_2 \to x'_1,y'_1 + 
x'_2,y'_2)\,
$$
\begin{equation} \label{ZEQADD}
u(x'_1,y'_1)\,u(x'_2,y'_2)\,\frac{1}{2} 
\frac{\partial}{\partial\,u(x_1,y_1)}\,\frac{\partial}{\partial\,u(x_2,y_2)}
\end{equation}
which will give the $2P \to P$ vertex in the form
\begin{equation} \label{V22}
\Gamma(2P \to P)(x_1,y_1 + x_2,y_2 \to x'_1,y'_1)\,=
\end{equation}
$$
 \,-\,\int d^2 x'_2\,d^2 
y'_2\,\,V_{2 \to 2}(x_1,y_1 + x_2,y_2 \to 
x'_1,y'_1 +
x'_2,y'_2)
$$
Therefore, we have to assume that $V_{2 \to 2} \,<\,0$. As we have discussed, 
generally speaking, it means that our problem has no solution. However, in QCD 
the situation is much better. Indeed, to generate  a correct  $2P \to P$ vertex
we need to introduce $V_{2 \to 2}\,\,\propto\,\bar{\alpha}^3_S$. The Feymann 
diagrams for this transition are obvious (see Fig. ~(\ref{feymdi}-b). However, the 
main contribution stems from the diagrams of Fig. ~(\ref{feymdi}-a) - type which 
are of the order of $\bar{\alpha}_S/N^2_c$. Therefore, diagrams of Fig. 
~(\ref{feymdi}-b) - type are small corrections to the main contribution and could 
be negative, in spite of the fact that the diagrams shown in  Fig.~(\ref{feymdi}-b)
actually give a positive contribution. We will discuss this scenario in 
all details in our paper \cite{L5}.

\section{Conclusions}

Being elegant and beautiful the BFKL Pomeron Calculus has a clear disadvantage: it 
is not a theoretically closed theory. Indeed, we need to add to the formalism of the 
BFKL Pomeron Calculus the theoretical ideas what kind of Pomeron interactions we 
should take into account and why.  Of course, the Feymann diagrams in leading log 
$1/x$ approximation of perturbative QCD allow us, in principle, to calculate all 
possible Pomeron 
interactions but, practically, it is a very hard job. Even if we will calculate these 
vertices we need to understand what set of vertices we should take into account 
for the 
calculation of the scattering amplitude.  This is the reason why we need to 
develop a more general formalism. Fortunately, such a formalism has been  
built and it is known under the abbreviation JIMWLK-Balitsky approach 
\cite{MV,JIMWLK,B}. In this approach we are able to calculate all vertices for 
Pomeron interactions as it was demonstrated in Ref. \cite{KOLU} and it solves the 
first part of the problem: determination of all possible Pomeron interactions. 
However, we need to understand what vertices we should take into account for the 
calculation of the scattering amplitude. We hope that a further progress in going
beyond of the BFKL Pomeron Calculus (see Refs. \cite{KOLU,HIMS}) will lead to such 
a development of the BFKL Pomeron Calculus that we will have a consistent 
theoretical 
approach. Hopefully this approach will be simpler than Lipatov's effective action 
\cite{LIEF} which is not easier to solve than the full QCD Lagrangian. 

The probabilistic interpretation gives a practical method for creating a Monte 
Carlo code in spirit of the approach suggested in Ref. \cite{MS}. This code will 
allow us to find a numerical solution to the problem and to consider the inclusive 
observables. This extension is very desirable since most 
of the experimental data 
exist just for these observables.

This year my teacher, Prof. Gribov, would have been 75 years old. I am happy to 
report him 
that his ideas are still alive and still give the simplest and the most elegant 
approach to high energy interaction. The high energy behaviour of the scattering 
amplitude is still an open theoretical question. We have learned a lot and we are 
able to foresee a lot of difficulties ahead.  Therefore, it is a good problem to be 
solved.

\end{document}